\documentclass[conference]{IEEEtran}
\IEEEoverridecommandlockouts
\usepackage{cite}
\usepackage{amsmath, amssymb, amsfonts, amsthm, bm, blkarray, comment, mathtools, enumitem}
\usepackage{algorithmic}
\usepackage{graphicx}
\usepackage{textcomp}
\def\BibTeX{{\rm B\kern-.05em{\sc i\kern-.025em b}\kern-.08em
    T\kern-.1667em\lower.7ex\hbox{E}\kern-.125emX}}

\usepackage[colorlinks=true, allcolors=blue]{hyperref}
\usepackage{booktabs}
\usepackage{lipsum}
\usepackage[table]{xcolor}
\usepackage{libertine}
\usepackage[subrefformat=parens]{subcaption}
\begin{document}
  
\title{Bubble Prediction of Non-Fungible Tokens (NFTs): An Empirical Investigation
}

\author{
\IEEEauthorblockN{Kensuke ITO}
\IEEEauthorblockA{\textit{\small Endowed Chair for Blockchain Innovation} \\
\textit{The University of Tokyo}\\
Tokyo, Japan \\
k-ito@g.ecc.u-tokyo.ac.jp}
\and
\IEEEauthorblockN{Kyohei SHIBANO}
\IEEEauthorblockA{\textit{\small Endowed Chair for Blockchain Innovation} \\
\textit{The University of Tokyo}\\
Tokyo, Japan \\
shibano@tmi.t.u-tokyo.ac.jp}
\and
\IEEEauthorblockN{Gento MOGI}
\IEEEauthorblockA{\textit{\small Endowed Chair for Blockchain Innovation} \\
\textit{The University of Tokyo}\\
Tokyo, Japan \\
mogi@tmi.t.u-tokyo.ac.jp}
}

\maketitle

\begin{abstract}
Our study empirically predicts the bubble of {\em non-fungible tokens} (NFTs): transferable and unique digital assets on public blockchains.
This topic is important because, despite their strong market growth in 2021, NFTs on a project basis have not been investigated in terms of bubble prediction.
Specifically, we applied the {\em logarithmic periodic power law} (LPPL) model to time-series price data associated with four major NFT projects.
The results indicate that, as of December 20, 2021, (i) NFTs, in general, are in a small bubble (a price decline is predicted), (ii) the {\em Decentraland} project is in a medium bubble (a price decline is predicted), and (iii) the {\em Ethereum Name Service} and {\em ArtBlocks} projects are in a small negative bubble (a price increase is predicted).
A future work will involve a prediction refinement considering the heterogeneity of NFTs, comparison with other methods, and the use of more enriched data.
\end{abstract}

\begin{IEEEkeywords}
non-fungible token, blockchain, bubble prediction, logarithmic periodic power law model
\end{IEEEkeywords}

\section{Introduction}

\begin{figure}[t]
\centering
 \includegraphics[width=0.9\hsize]{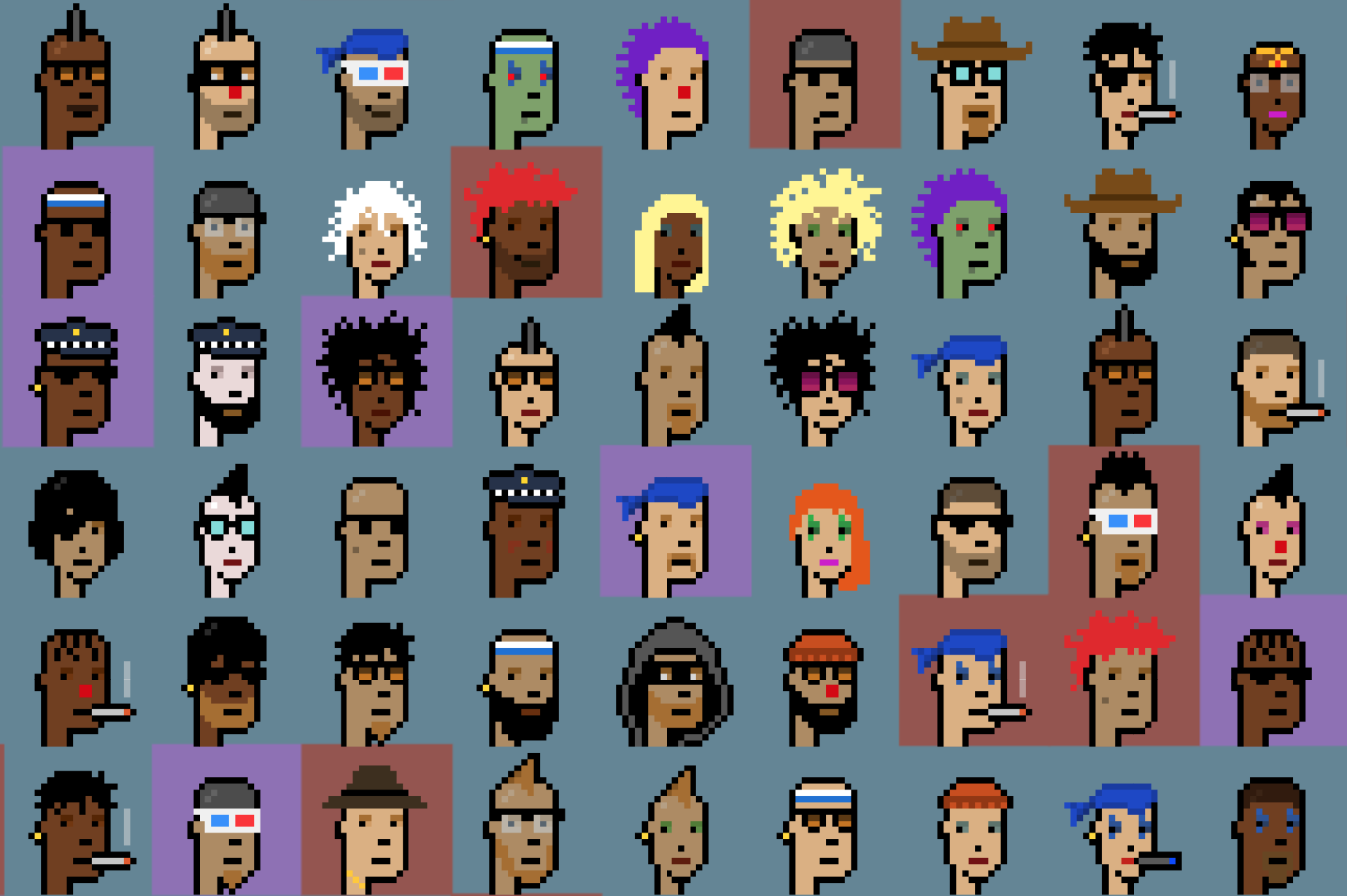}
 \captionsetup{font=footnotesize}
    \caption*{{\em Source}: CryptoPunks (\url{https://www.larvalabs.com/cryptopunks}, accessed November 24, 2021).}
    \captionsetup{font=normal}
 \caption{Several NFTs Minted by CryptoPunks}
 \label{punks}
\end{figure}

{\em Non-fungible tokens} (NFTs) are ``transferable and unique digital assets on public blockchains" \cite{ante2021nona}\footnote{Other definitions may not limit NFTs to those used in the public blockchain, or conversely, may limit NFTs to those used in the Ethereum \cite{buterin2014next, wood2014ethereum} public blockchain.
For example, ``An NFT is a unit of data stored on a blockchain that certifies a digital asset to be unique and therefore not interchangeable" \cite{nadini2021mapping2}, ``Non-Fungible Token (NFT) is a type of cryptocurrency that is derived by the smart contracts of Ethereum" \cite{wang2021non}.} developed as an extension to cryptocurrencies (e.g., Bitcoin; BTC \cite{nakamoto2008bitcoin}), enabling decentralized consensus-building on transaction records. 
Unlike fungible (homogeneous) cryptocurrencies\footnote{Note that several studies (e.g., \cite{meiklejohn2015privacy, conti2018survey, amarasinghe2019survey}) have pointed out that cryptocurrencies, where all transaction records are public, are non-fungible because their values can vary between those that went through a particular address and those that did not.}, NFTs can be unique digital assets (e.g., arts, games, collectibles, etc.), once they are associated with unique metadata or images \cite{wang2021non, franceschet2021crypto}.
In practice, NFTs are often minted on a {\em project} basis as a group of similar images.
For example, {\em CryptoPunks}\textemdash one of the earliest NFT projects\textemdash minted 10,000 NFTs, each with a human face drawn on a 24 $\times$ 24 pixel image (Fig. \ref{punks}).
According to \href{https://coinmarketcap.com/nft/collections/}{coinmarketcap.com}, there are currently at least 1,056 such projects on Ethereum (ETH)\cite{buterin2014next, wood2014ethereum}\footnote{If other blockchains (e.g., Binance Smart Chain and Solana) are included, the number of projects increases to 1,998.}.

\begin{table*}[t]
 \caption{Five Most Expensive NFTs (as of December 20, 2021)}
 \label{nftrank}
 \begin{center}
 \rowcolors{1}{white}{gray!10}
  \begin{tabular}{lllll}
   \toprule
   \textbf{Rank} & \textbf{Title} & \textbf{Artist} & \textbf{Price (USD)} & \textbf{Date}  \\[2pt]
   \midrule
1 & {\em EVERYDAYS: THE FIRST 5000 DAYS} & beeple & \$$69,346,250.00$ & March 11, 2021\\[2pt]
2  & {\em HUMAN ONE} & beeple & \$$28,985,000.00$ & November 10, 2021\\[2pt]
3 & {\em Stay Free (Edward Snowden, 2021)} & snowden & \$$9,516,829.60$ & April 16, 2021\\[2pt]
4  & {\em CROSSROAD} & beeple & \$$6,600,000.00$ & February 25, 2021\\[2pt]
5  & {\em OCEAN FRONT} (beeple) & beeple & \$$6,000,000.00$ & March 20, 2021\\[2pt]
   \bottomrule
  \end{tabular}
   \captionsetup{font=footnotesize}
    \caption*{{\em Source}: MOST EXPENSIVE ARTWORKS (\url{https://cryptoart.io/}, accessed December 20, 2021), created by the author.}
 \end{center}
\end{table*}

\begin{figure}[t]
\centering
 \includegraphics[width=1.0\hsize]{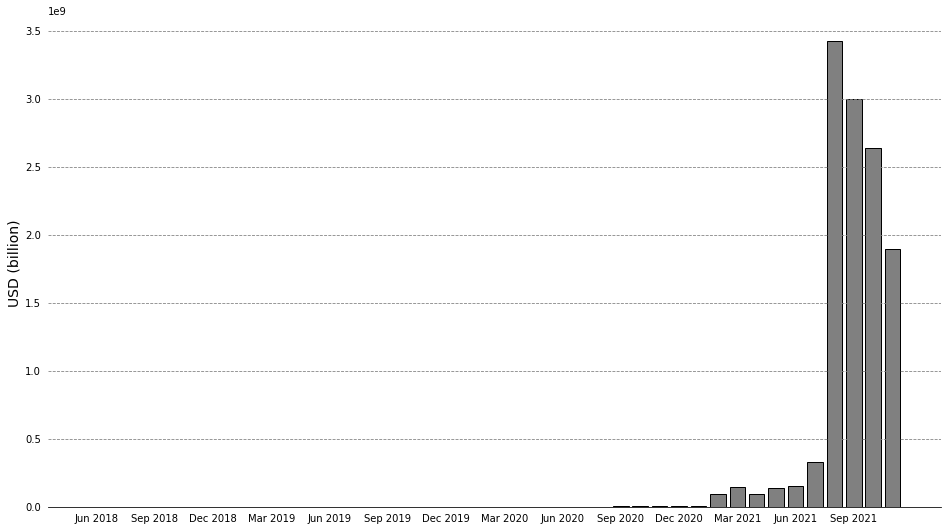}
 \captionsetup{font=footnotesize}
    \caption*{{\em Source}: @rchen8 / OpenSea monthly volume (\url{https://dune.xyz/queries/3469/6913}, accessed November 24, 2021), created by the author.}
    \captionsetup{font=normal}
 \caption{Monthly Volume of NFTs Traded in OpenSea}
 \label{market}
\end{figure}

One of the most notable features of NFTs (at the time of writing this paper) is their strong market growth.
Fig. \ref{market} presents the monthly volume of NFTs traded in the world's largest NFT marketplace {\em OpenSea}. 
We can clearly observe a huge market growth trend since 2021, especially in August.
The same trend has also been observed in luxury NFTs traded outside the {\em OpenSea}.
Table \ref{nftrank} shows the ranking of the most expensive NFTs traded on seven marketplaces ({\em SuperRare}, {\em Nifty Gateway}, {\em Foundation}, {\em hic et nunc}, {\em MakersPlace}, {\em KnownOrigin}, and {\em Async Art}), where the prices of the top five NFTs were all recorded in 2021.
These trends of strong market growth naturally lead people to question whether the current NFT market is a bubble \cite{howcroft2021nft, castano2021nft} or what the future price movements will be \cite{zoupnon2021sorry, canny2022jeff}.

To answer these questions, this study attempts a bubble prediction of NFTs, especially for each project.
Specifically, we applied the {\em logarithmic periodic power law} (LPPL) model \cite{johansen1999predicting, johansen2000crashes, sornette2009stock}\textemdash a standard method for bubble prediction\textemdash to time-series price data associated with four major NFT projects.
These data were retrieved from \href{https://nonfungible.com/}{nonfungible.com} (see Section \ref{data} for details).
The results imply that, as of December 20, 2021, (i) NFTs, in general, are in a small bubble (a price decline is predicted), (ii) the {\em Decentraland} project is in a medium bubble (a price decline is predicted), and (iii) the {\em Ethereum Name Service} and {\em ArtBlocks} projects are in a small negative bubble (a price increase is predicted).
The analysis presented here is relatively intuitive and, to the best of the authors' knowledge, is the first empirical investigation focusing on the bubble prediction of NFT projects.

This paper consists of six sections, including the introduction.
Section \ref{related} covers the related research work; 
Section \ref{data} details the retrieved time-series price data;
Section \ref{methods} describes the LPPL model for the bubble prediction; 
Section \ref{results} shows the results and their implications;
finally, Section \ref{conclusion} concludes this study and outlines future research directions.

\section{Related Research Work} \label{related}

\subsection{Empirical Study on NFT Markets}
Thanks to the transparency of transaction records, empirical studies on NFT markets, using richer data (both in quality and quantity), are now emerging.

Nadini, et al. \cite{nadini2021mapping2} presented the first empirical study on NFTs.
In this study, massive transaction data (6.1 million NFT trades between June 23, 2017 and April 27, 2021) were comprehensively analyzed using a variety of methods, including network analysis for community detection, neural networks for image classification, and linear regression for price estimation. 
Ante \cite{ante2021nona} specifically focused on the spillover effect among NFT projects and found a Granger causality between the number of active wallets in established projects and that in emerging projects\footnote{Ante \cite{ante2021nona} found that the decreasing number of active wallets in emerging projects causes an increase in the number of active wallets in established projects and vice versa.}.
Moreover, Ante \cite{ante2021nonb} and Dowling \cite{dowling2021fertile, dowling2021non} investigated the spillover between NFTs and cryptocurrecies.
The former author found Granger causalities from BTC price to NFT sales, from ETH price to the number of active NFT wallets, and from BTC price to ETH price.
On the other hand, the latter author pointed out that there is little causality between NFTs and cryptocurrecies with respect to price volatilities.

Regarding this research topic, our study has an academic significance in terms of covering the bubble prediction, while focusing on NFT markets.

\subsection{Bubble Prediction on Cryptocurrencies}
Bubble prediction on cryptocurrencies, especially BTC, has been a popular topic since we can directly apply preceding techniques used in the traditional financial markets.

To the best of the authors' knowledge, Macdonell \cite{macdonell2014popping} presented the first study on this topic and applied the LPPL model to weekly moving-average BTC prices (from July 2010 to August 2013), thereby predicting the bubble crash in December 2013.
Subsequent studies have extended this approach in various ways, such as applying the LPPL model to other cryptocurrencies \cite{rokosz2018evaluating, bianchetti2018cryptocurrencies}, adding new terms into the model \cite{wheatley2019bitcoin, xiong2020new}, and modifying the model itself to accommodate highly-volatile BTC price data \cite{shu2020real}.
The LPPL model is the standard, but not the only method for bubble prediction.
Preceding studies have also adopted other methods, including the augmented Dickey-Fuller test \cite{cheung2015crypto, corbet2018datestamping, wheatley2019bitcoin}, sentiment analysis \cite{bukovina2016sentiment, karalevicius2018using, chen2019sentiment}, and machine learning \cite{mallqui2019predicting, chen2020bitcoin, khedr2021cryptocurrency}\footnote{For more information, see Kyriazis et al. \cite{kyriazis2020systematic}, a survey paper that covers not only predictions, but the entire phenomenon of cryptocurrency bubbles.}.

Regarding this research topic, our study has an academic significance in terms of covering NFT markets, while focusing on the bubble prediction.

\subsection{Bubble Prediction on DeFi and NFTs}
In parallel with our study, the scope of bubble prediction is being extended from cryptocurrencies to decentralized finance (DeFi) \cite{schar2021decentralized} and NFTs.

Maouchi et al. \cite{maouchi2021understanding} presented the first study on this topic and pointed out fundamental properties of DeFi and NFT bubbles (e.g., they overlap with cryptocurrency bubbles with the exception of summer 2020; they are less recurrent but larger than cryptocurrency bubbles).
Wang et al. \cite{wang2022bubbles} elaborated this extension by using the augmented Dickey-Fuller test and (supplementally) the LPPL model.

Regarding this research topic, our study has an academic significance in terms of providing bubble prediction for each project, while focusing on the NFT market.

\section{Data} \label{data}

\begin{figure*}[t]
\centering
 \includegraphics[width=0.9\hsize]{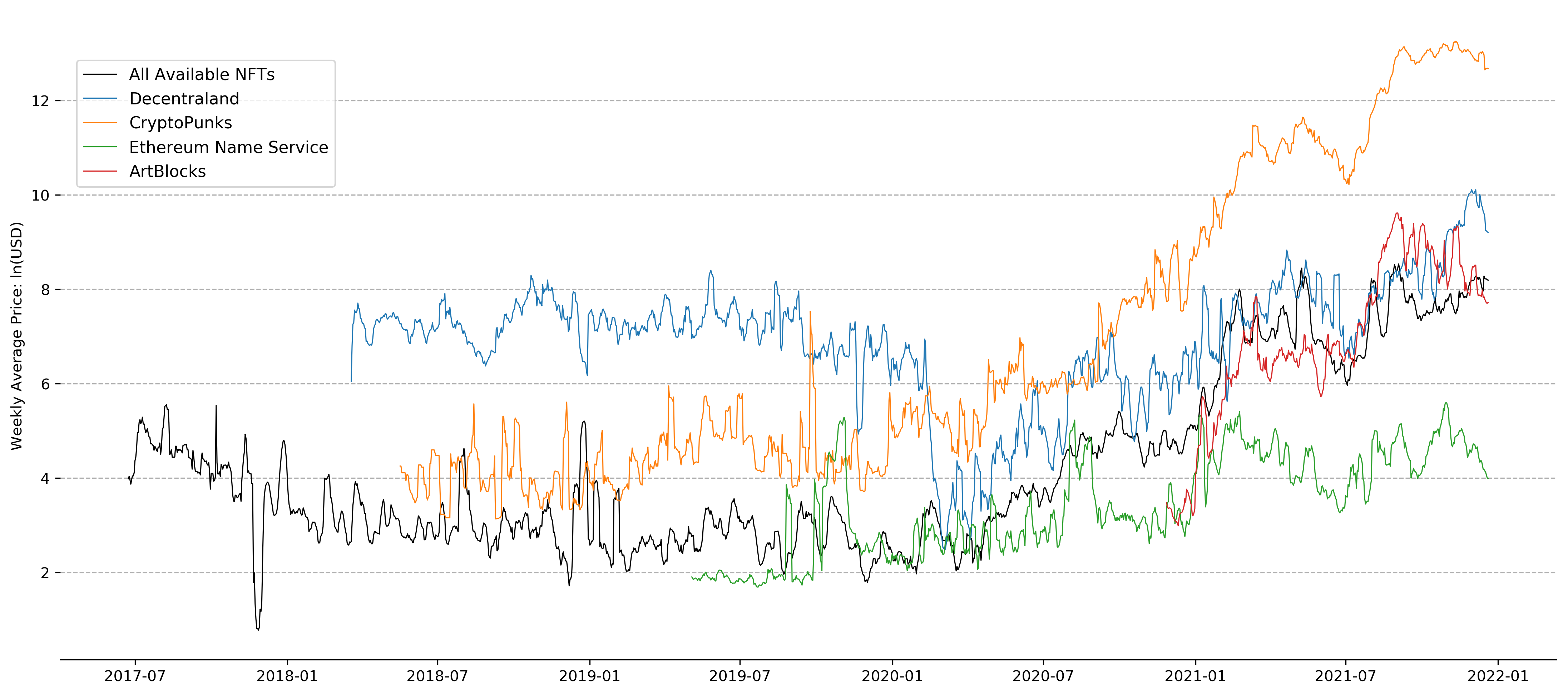}
 \caption{The Time-Series Price Data Used in Our Analysis}
 \label{total}
\end{figure*}

The time-series data used in this analysis are the weekly moving-average prices of NFTs, which were retrieved from \href{https://nonfungible.com/}{nonfungible.com} and displayed on a daily basis in US dollars.
The moving-average prices were extracted not only from all available NFTs\footnote{According to \href{https://nonfungible.com/}{nonfungible.com}, over 8 million sales records from more than 200 NFT projects have been tracked.} (from 2017-06-23 to 2021-12-20), but also from each of the four major projects with different time-scales and categories:
{\em Decentraland} (from 2018-03-19 to 2021-12-20),
{\em CryptoPunks} (from 2018-05-17 to 2021-12-20),
{\em Ethereum Name Service} (from 2019-05-04 to 2021-12-20), and
{\em ArtBlocks} (from 2020-11-27 to 2021-12-20)\footnote{All data are available from the \href{https://github.com/knskito/NFTempirical}{Github repository}.}. 
Fig. \ref{total} plots the retrieved data using log-scale values on the vertical axis.
Our analysis applies the LPPL model to each of these 1 + 4 time-series price data.

The following simplifications are considered: (i) weekly moving-average price data are used (instead of daily moving-average data) to facilitate the application of the LPPL model to the highly-volatile NFT market\footnote{Macdonell \cite{macdonell2014popping} used weekly moving-average BTC price data, whereas daily moving-average values were used in subsequent studies \cite{rokosz2018evaluating, bianchetti2018cryptocurrencies, wheatley2019bitcoin, xiong2020new, shu2020real}.};
(ii) price data associated with projects (or all available NFTs) are used (instead of those associated with each NFT) to facilitate the application of the LPPL model to the NFT market where each one has a unique price\footnote{The low liquidity this feature brings is one of the reasons for the NFT market high volatility.}.
In other words, it implicitly assumes that all NFTs are homogeneous for each project.
Relaxing these two simplifications (which undermine the original characteristics of the NFT market) is one of our future research directions.

\section{Methods} \label{methods}

\subsection{LPPL Model}
The LPPL model is designed to predict bubbles using only time-series price data.
Specifically, it approximates $\ln{[p(t)]}$\textemdash the log-price of data at a given period $t$\textemdash as follows:

\begin{equation} \label{lppl}
\begin{split}
  \ln{[p(t)]} \approx A + B(t_{c} - t)^{m} + C(t_{c} - t)^{m}\\ 
  \cdot\cos[\omega\ln{(t_{c} - t)} - \phi],
\end{split}
\end{equation}

\noindent
where the right-hand side contains three linear parameters ($A, B,$ and $C$) and four nonlinear parameters ($t_c, m, \omega,$ and $\phi$).

$t_c$ denotes the {\em critical time} at which the previous bubble ends and a transition to another regime occurs\footnote{The critical time is also called singularity. This is the reason the LPPL model is also abbreviated as LPPLS model.}; 
$A$ denotes the log-price when $t$ reaches $t_c$;
$B$ denotes the magnitude of the power law acceleration\footnote{Note that the power law acceleration works toward increasing prices if $B < 0$ and toward decreasing prices if $B > 0$. The former case is denoted as a {\em positive bubble}, whereas the latter case is denoted as a {\em negative bubble} \cite{yan2010diagnosis}.};
$C$ denotes the magnitude of the log-periodic oscillations ($|C| < 1$);
$m$ denotes the degree of the power law acceleration;
$\omega$ denotes the frequency of oscillations during the bubble;
$\phi$ denotes the time scale of oscillations.

Namely, the LPPL model defines a bubble as the {\em faster-than-exponential} (upward or downward) acceleration of $p(t)$.

\subsection{Calibration}

The LPPL model can be calibrated using the ordinary least squares method to provide estimations for all parameters ($A, B, C, t_c, m, \omega,$ and $\phi$) in a given time window $[t_1, t_2]$.
We here employed an existing \href{https://pypi.org/project/lppls/#description}{Python module} for this calibration and visualization.

To reduce the computational complexity, the calibration initially uses the Filimonov and Sornette \cite{filimonov2013stable} method, which eliminates the nonlinear parameter $\phi$ by expanding Equation \ref{lppl} as follows:

\begin{equation} \label{lpplfit}
\begin{split}
  \ln{[p(t)]} \approx A + (t_{c} - t)^{m}\{B + C_{1}\cos{[\omega\ln{(t_{c} - t)}]} \\
  + C_{2}\sin{[\omega\ln{(t_{c} - t)}]}\},
\end{split}
\end{equation}

\noindent
where $C_{1} = C\cos\phi$ and $C_{2} = C\sin\phi$\footnote{Since $|C| < 1$, condition $C_{1}^{2} + C_{2}^{2} < 1$ holds.}.

For the remaining nonlinear parameters $t_c, m,$ and $\omega$, we set the following conditions, which are derived from the empirical evidence of previous bubbles and are commonly adopted as the stylized features of the LPPL model \cite{sornette2001significance, johansen2010shocks, sornette2015real}:

\begin{align}\label{nonlinear}
    \max\left\{t_2 - 60, \frac{t_2 - 0.5}{t_2 - t_1}\right\} &< t_{c} < \min\left\{t_2 + 252, \frac{t_2 + 0.5}{t_2 - t_1}\right\},\\[2pt]
    0 &< m < 1,\\[2pt]
    2 &< \omega < 15.
\end{align}

\noindent
The calibration for a given time window $[t_1, t_2]$ ends if the estimated parameters satisfy all the above stylized features\footnote{Accordingly, this calibration is stochastic rather than deterministic (i.e., the calibration results are not unique and vary in each iteration).}.

Then, bubble prediction can be made for each data period by letting this calibration iterate for the shrinking time window $[t_1, t_2]$.
Here $[t_1, t_2]$ is in daily units, according to the daily basis data (Section \ref{data}); $t_2$ denotes a fictitious today corresponding to $t$, and
$t_1$ denotes an earlier day.
For a given $t_2$, the iterative calibration sets the initial range of the time window to 120 days and the shrinking interval of $t_1$ to 5 days.
Thus, the parameters need to be estimated 24 times for each $t_2$ (e.g., [1, 120], [5, 120], $\cdots$ , [115, 120] for $t_2=120$; [2, 121], [6, 121], $\cdots$ , [116, 121] for $t_2=121$)\footnote{Thus, this analysis exhibits a lag for approximately four months; the results for all available NFTs (from 2017-06-23 to 2021-12-20) range from 2017-10-20 to 2021-12-20.}. 
It should be emphasized that this process (as a prediction) uses only historical data as input data.
The outcome of $t_2$ depends only on data from $t_2$ to the last 120 days.

\subsection{Bubble Indicator for Visualization}

The LPPL model visualizes its own predictions as the {\em bubble indicator} (or {\em confidence indicator}) \cite{sornette2015real}.
For a given $t_2$ with 24 outcomes, the bubble indicator first counts the number of $B<0$ and $B>0$ cases; the former implies that the price increases faster than exponential (i.e., positive bubble), whereas the latter implies that the price decreases faster than exponential (i.e., negative bubble \cite{yan2010diagnosis}).
These numbers are denoted as $[B < 0]_{count}$ and $[B > 0]_{count}$ for convenience.

Next, the 24 outcomes are classified into $B<0$ and $B>0$ groups and filtered to obtain a higher degree of confidence (thereby preventing the type I error).
This process specifically sets the following {\em filter conditions} for the nonlinear parameters $t_c, m,$ and $\omega$: 

\begin{align}
    \frac{\omega}{2\pi}\cdot\ln\frac{t_c - t_1}{t_c - t_2} & > 2.5,\label{filter}\\[4pt]
    \frac{m|B|}{\omega|C|} & > 0.5,\label{filter2}
\end{align}

\noindent
where the number of outcomes that satisfy the above filter conditions in the $B<0$ and $B>0$ groups are denoted as $[B < 0]^{*}_{count}$ and $[B > 0]^{*}_{count}$, respectively.

Finally, the bubble indicators (for a given $t_2$) can be computed as follows:

\begin{align}
    bubble indicator(pos) = \frac{[B < 0]^{*}_{count}}{[B < 0]_{count}},\\[4pt]
    bubble indicator(neg) = \frac{[B > 0]^{*}_{count}}{[B > 0]_{count}},
\end{align}

\noindent
where $bubbleindicator(pos)$ indicates the percentage of a positive bubble of the price at $t_2$ in the $[0, 1]$ range, which quantifies the possibility of a price decline in the near future.
On the other hand, $bubbleindicator(neg)$ indicates the percentage of a negative bubble of the price at $t_2$ in the $[0, 1]$ range, which quantifies the possibility of a price increase in the near future\footnote{Here, the bubble indicator is regarded as zero if the denominator ($[B < 0]_{count}$ or $[B > 0]_{count}$) is zero.}.

The LPPL model derives these positive and negative bubble indicators for all $t_2$, thereby visualizing its own predictions.

\section{Results} \label{results}

\begin{figure*}[!t]
 \begin{minipage}[!t]{1.0\hsize}
 \centering
 \includegraphics[width=0.9\hsize]{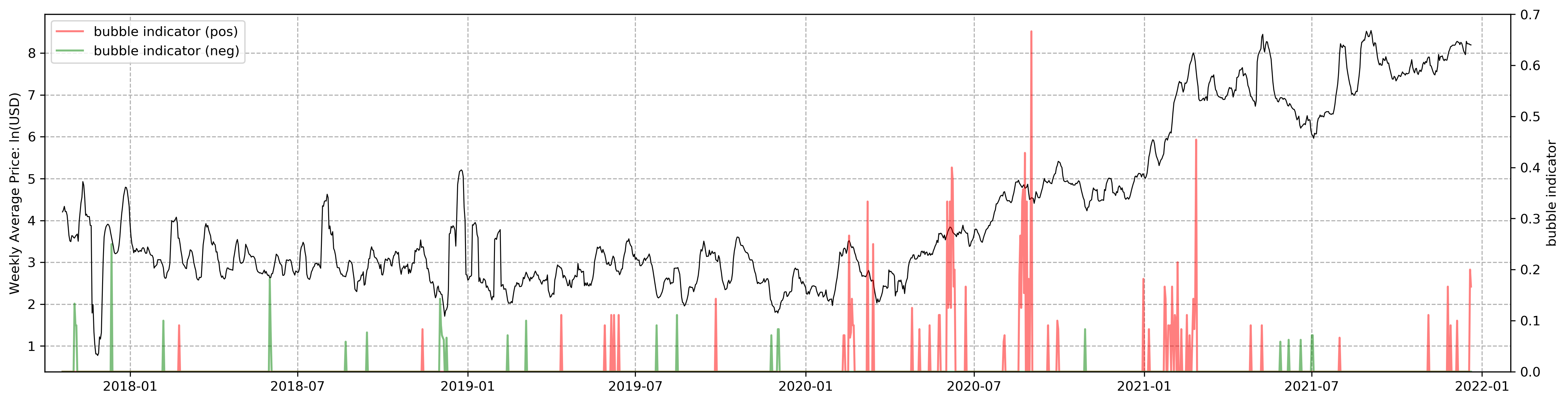}
 \vspace{-5pt}
 \captionsetup{labelfont=small}
    \subcaption{All Available NFTs} \label{resulta}
 \vspace{5pt}
 \end{minipage}
 \begin{minipage}[!t]{1.0\hsize}
 \centering
 \includegraphics[width=0.9\hsize]{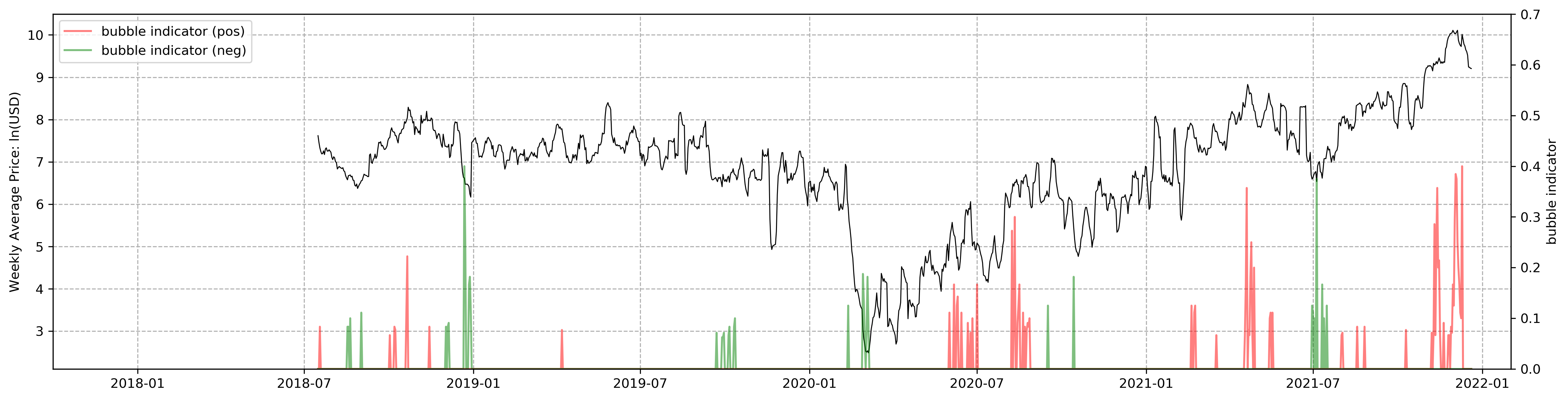}
 \vspace{-5pt}
 \captionsetup{labelfont=small}
    \subcaption{Decentraland} \label{resultb}
 \vspace{5pt}
 \end{minipage}
 \begin{minipage}[!t]{1.0\hsize}
 \centering
 \includegraphics[width=0.9\hsize]{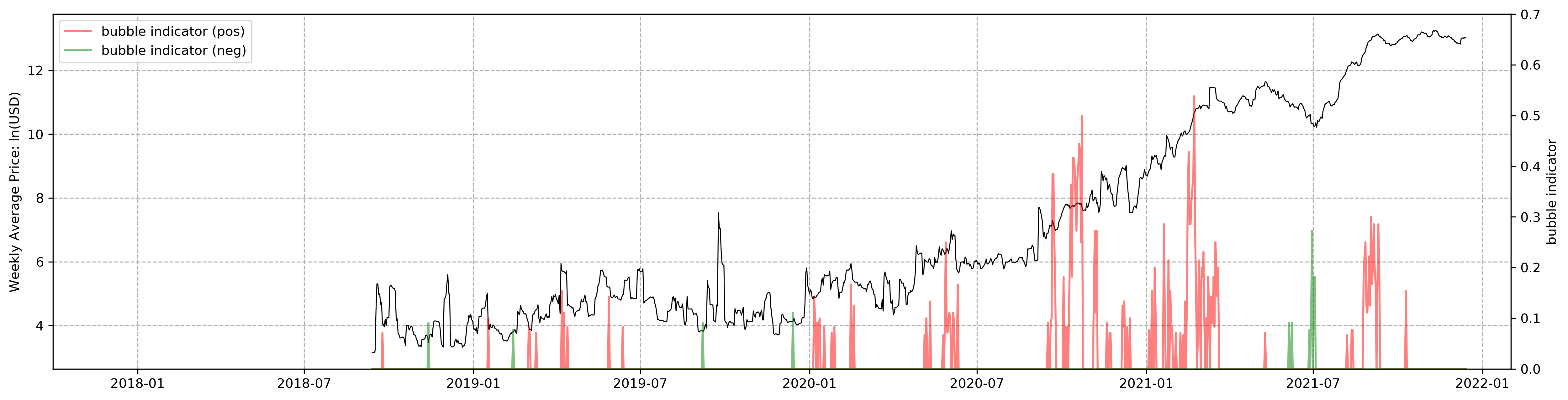}
 \vspace{-5pt}
 \captionsetup{labelfont=small}
    \subcaption{CryptoPunks} \label{resultc}
 \vspace{5pt}
 \end{minipage}
 \begin{minipage}[!t]{1.0\hsize}
 \centering
 \includegraphics[width=0.9\hsize]{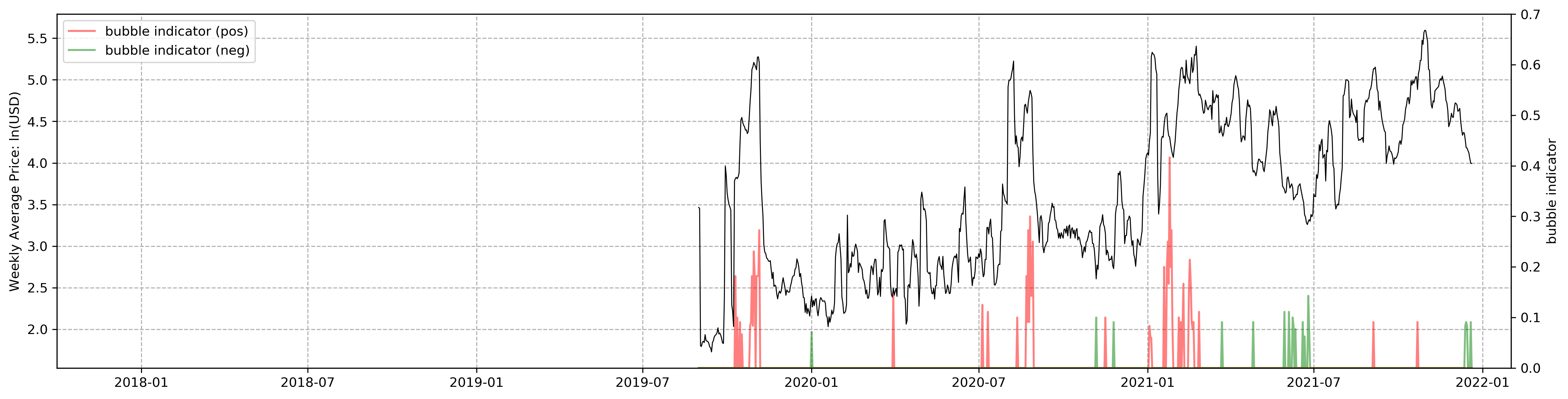}
 \vspace{-5pt}
 \captionsetup{labelfont=small}
    \subcaption{Ethereum Name Service} \label{resultd}
 \vspace{5pt}
 \end{minipage}
 \begin{minipage}[!t]{1.0\hsize}
 \centering
 \includegraphics[width=0.9\hsize]{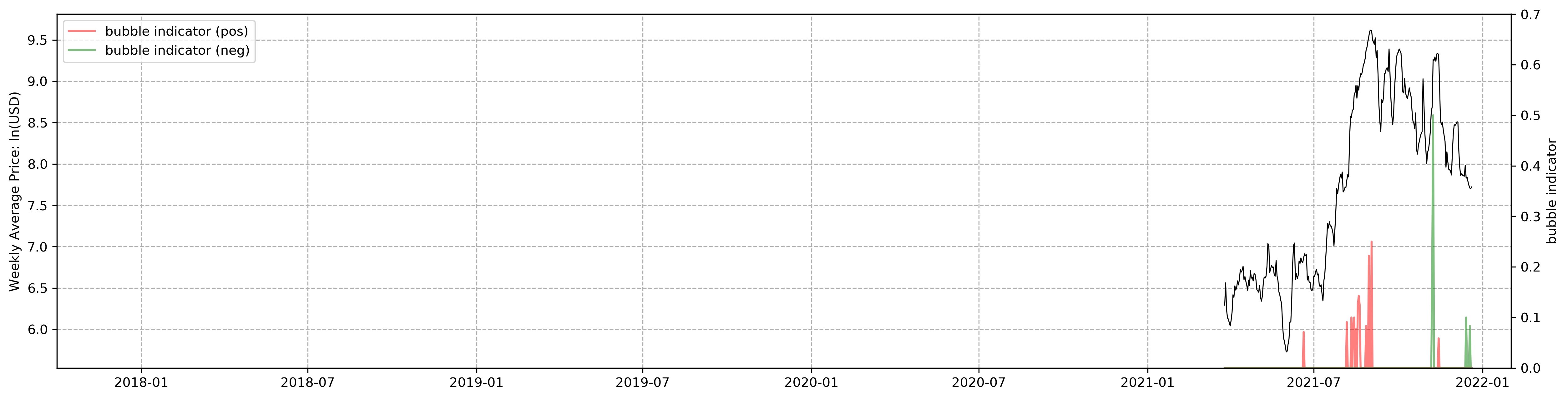}
 \vspace{-5pt}
 \captionsetup{labelfont=small}
    \subcaption{ArtBlocks} \label{resulte}
 \end{minipage}
 \caption{Prediction Results}
 \label{figresults}
\end{figure*}

The prediction results are summarized in Fig. \ref{figresults}.
Figs. \ref{resulta}-\ref{resulte} correspond to the aforementioned 1 + 4 time-series price data, where $bubbleindicator(pos)$ is depicted in red and $bubbleindicator(neg)$ in green.

Overall, the LPPL model appears to predict the trend of both positive and negative bubbles in NFT markets.
The results obtained considering all the available NFTs (Fig. \ref{resulta}) show that the LPPL model is generally successful in predicting the direction of price changes, although prices rise further after $bubbleindicator(pos)$ reaches its highest value in late August 2020.
This is true for individual projects as well; $bubbleindicator(neg)$ successfully predicts the shift to an upward price trend after July 2021, which is common across the {\em Decentraland} (Fig. \ref{resultb}), {\em CryptoPunks} (Fig. \ref{resultc}), and {\em Ethereum Name Service} (Fig. \ref{resultd}), although $bubbleindicator(pos)$ fails to predict the continuous price increase of the {\em CryptoPunks} (Fig. \ref{resultc}) from around October 2020 to March 2021. 

After confirming the accuracy of the LPPL model, let us focus on the latest indicators.
Near December 20, 2021, the bubble indicators provide signals in four cases, except for {\em CryptoPunks} (Fig. \ref{resultc}); $bubbleindicator(pos) \approx 0.2$ in all available NFTs (Fig. \ref{resulta}), $bubbleindicator(pos) \approx 0.4$ in the {\em Decentraland} (Fig. \ref{resultb}), and $bubbleindicator(neg) \approx 0.1$ in the {\em Ethereum Name Service} (Fig. \ref{resultd}) and {\em ArtBlocks} (Fig. \ref{resulte}).
These results imply that, as of December 20, 2021, (i) NFTs, in general, are in a small bubble (a price decline is predicted), (ii) the {\em Decentraland} project is in a medium bubble (a price decline is predicted), and (iii) the {\em Ethereum Name Service} and {\em ArtBlocks} projects are in a small negative bubble (a price increase is predicted).

\section{Conclusion} \label{conclusion}
In this paper, we empirically predicted the bubble of NFTs, by applying the LPPL model to the time-series price data associated with four major NFT projects.
The results indicated that, as of December 20, 2021, (i) NFTs, in general, are in a small bubble (a price decline was predicted), (ii) the {\em Decentraland} project is in a medium bubble (a price decline was predicted), and (iii) the {\em Ethereum Name Service} and {\em ArtBlocks} projects are in a small negative bubble (a price increase was predicted).
To the best of the authors' knowledge, this is the first empirical investigation focusing on the bubble prediction of NFT projects.

On the other hand, this study, as a first step of the prediction, needs further investigation to obtain more accurate results.
The following three potential research directions are outlined:

\subsection{Consideration of the NFTs Heterogeneity}
We assumed that, to facilitate the application of the LPPL model, all NFTs are homogeneous for each project (Section \ref{data}). 
However, NFTs are inherently unique and heterogeneous, and this is exactly what differentiates NFTs from cryptocurrencies (and legal tender).
Therefore, it is worth addressing to develop a new prediction method which will take into account the heterogeneity of NFTs.

\subsection{Comparison with Other Methods}
The LPPL model is only one of the available methods for bubble prediction.
Other methods, such as the augmented Dickey-Fuller test, the sentiment analysis, and machine learning (Section \ref{related}), can also be used. 
Using these methods and comparing their results would be another future research direction.
Moreover, we need to find the {\em best mix} of the preceding methods to achieve the best accuracy of NFT bubble prediction.
In this case, a variety of data (other than time-series price) would be required as input data.

\subsection{Use of More Enriched Data}
Our prediction deals with only four major NFT projects, although it uses the weekly moving-average prices of all available NFTs (Section \ref{data}).
More enriched data obtained from other projects would refine the analysis.
In addition, although weekly moving-average prices (to address the high volatility) were used, it would be important to develop new methods that can make use of daily or hourly time-series price data as input data.



\section*{Acknowledgment}

The authors would like to express their gratitude to the \href{https://nonfungible.com/}{nonfungible.com} for providing the time-series price data of NFTs.

\bibliography{refs} 
\bibliographystyle{IEEEtran} 

\end{document}